\documentclass[nofootinbib,prd,superscriptaddress]{revtex4}

\usepackage{graphicx}
\usepackage{mathrsfs}
\usepackage{yfonts}
\usepackage{tipa}
\usepackage{bm}
\usepackage{bbm}
\usepackage{amsmath}
\usepackage{amssymb}
\usepackage{amsthm}
\usepackage{hyperref}
\usepackage{amsfonts}
\usepackage{mathtools}
\usepackage{latexsym}
\usepackage[greek,english]{babel}
\usepackage{booktabs}
\usepackage{xcolor}
\usepackage[iso-8859-7]{inputenc}

\begin{document}

\title{Warp Drive in a De Sitter Universe}

\author{Remo Garattini}
\email{remo.garattini@unibg.it}
\affiliation{Universit\`a degli Studi di Bergamo, Dipartimento di Ingegneria e Scienze
	Applicate, Viale Marconi 5, 24044 Dalmine (Bergamo) Italy and I.N.F.N.-
	sezione di Milano, Milan, Italy.}
\author{Kirill Zatrimaylov}
\email{kirill.zatrimaylov@unibg.it}
\affiliation{Universit\`a degli Studi di Bergamo, Dipartimento di Ingegneria e Scienze
	Applicate, Viale Marconi 5, 24044 Dalmine (Bergamo) Italy.}

\begin{abstract}
Generalizing the result of H. Ellis who embedded a warp bubble in the background of a black hole, we introduce a warp bubble in a de Sitter universe. We show that under certain conditions (namely, that the bubble is moving in the radial direction at a velocity equal to the speed of the expansion of the universe), it is possible for the bubble to have strictly non--negative energy density, with the weak and null energy satisfied up to a total divergence term that averages to zero. We discuss the implications of this result and its possible applications to models of dark energy like "dark fluid" and quintessence, as well as to physical systems like Casimir cavities and analogue gravity setups.
\end{abstract}
\maketitle
\section{Introduction}\label{introduction}
The concept of warp drives was first introduced by Miguel Alcubierre in his seminal 1994 paper~\cite{Alcubierre:1994tu}, and elaborated upon by Jose Natario in~\cite{Natario:2001tk}. A warp drive is a solution of General Relativity that has the appearance of a "bubble" propagating on some (flat or non--flat) spacetime background. The observers inside the bubble are in an inertial reference frame, which means warp drives do not require external energy sources to accelerate, and they may move at any speed (in principle including superluminal). This makes them a viable candidate for interstellar travel, but they have one significant downside: in order to sustain a bubble, one requires exotic matter with negative energy density~\cite{Natario:2001tk}--\cite{Alcubierre:2017kqf}.

As pointed out by Homer Ellis in~\cite{Ellis:2004aw}, the Schwarzschild metric, which describes a black hole, can be mapped to a warp drive--type metric with the use of a coordinate system known as Painleve--Gullstrand coordinates, which makes it possible to embed a warp drive in a black hole background. In~\cite{Garattini:2024zpz}, we studied the implications of this result, and in~\cite{Garattini:2024pgk}, we generalized it to wormholes. In this paper, we generalize this result to the case of a warp bubble in a de Sitter universe and demonstrate that it's possible for it to have a non--negative energy density, meaning that it may be physically interpreted not as an object made of some exotic matter, but rather as a vacuum energy inhomogeneity. This kind of object satisfies the null and weak energy conditions up to a total divergence term that averages to zero.

First, we briefly outline the warp drive physics in general and Ellis' formalism in section~\ref{sec:Intro}. Then, in section~\ref{sec:DeSitter}, we apply this formalism to the de Sitter metric. We conclude in section~\ref{sec:conclusions} with an overview of these results and their possible implications for cosmology and lab setups.

\section{An Overview of Warp Drives}\label{sec:Intro}
A warp drive metric, as defined by Alcubierre in~\cite{Alcubierre:1994tu}, is a metric of the form
\begin{equation}\label{Alc}
-dt^2+(dx-fvdt)^2+dy^2+dz^2 \ ,
\end{equation}
where $f(r_s)$ is a function that is equal to 1 for $r_s$ smaller than some scale $R_W$, and quickly decreases to zero for $r_s>R_W$. Here
\begin{equation}
r_s=\sqrt{(x-x_0(t))^2+y^2+z^2} \ ,
\end{equation}
and
\begin{equation}
v \ = \ \frac{dx_0}{dt} \ .
\end{equation}
Physically, this metric describes a localized "bubble" with a radius $R_W$, whose center is located at $\left(x_0(t),0,0\right)$, and within which the observer is in a reference frame moving at velocity $v$ (which is inertial if $v=const$). The Alcubierre metric belongs to a more generic class of metrics of the form
\begin{equation}
-dt^2+\sum^3_{i=1}\left(dx^i+N^idt\right)^2 \ ,
\end{equation}
known as the Natario class~\cite{Natario:2001tk}. In ADM terms, this is a type of metric that has the lapse $N=1$ and flat intrinsic metric ($h_{ij}=\delta_{ij}$). For every spacetime belonging to this class, it is possible to embed a warp bubble into it by changing the shift vector in the following way~\cite{Ellis:2004aw}:
\begin{equation}
N^i \ \rightarrow \ (1-f)N^i-fv^i \ .
\end{equation}
As shown in~\cite{Santiago:2021aup}, for a Natario--type metric, the Eulerian energy density is given by
\begin{equation}
\rho \ = \ \frac{1}{16\pi G}\left(\partial_i(N_i\partial_jN_j-N_j\partial_jN_i)-\frac{1}{4}(\partial_iN_j-\partial_jN_i)^2\right) \ ,
\end{equation}
where the first term is a total divergence that averages to zero for asymptotically flat spacetimes with $N^i$'s falling off quicker than $r^{-1/2}$ (and, hence, has to be negative \textit{somewhere}), and the second term is negative--definite. This means that asymptotically flat Natario spacetimes \textit{always} have regions of negative energy density; in particular, for the Alcubierre metric~\eqref{Alc}, the energy density is manifestly negative:
\begin{equation}
\rho \ = \ -\frac{f'^2v^2}{32\pi G}\left(\frac{y^2+z^2}{r^2_s}\right) \ .
\end{equation}
However, every theorem is only as strong as its assumptions: namely, if one takes a shift vector that is \textit{both} irrotational (i. e. it can be written as a gradient of some function $\phi$, as was also done in~\cite{Lentz:2021rzh,Fell:2021wak}) \textit{and} asymptotically non-vanishing (or falling off slower than $r^{-1/2}$), it's possible to produce $\rho$ that is non--negative everywhere. By combining this result with the Ellis' procedure of "embedding", we demonstrate the possibility of non--negative energy density for a warp bubble embedded in a de Sitter universe.
\section{De Sitter Spacetime}\label{sec:DeSitter}
Beginning with the standard FRW form of the de Sitter metric:
\begin{equation}
-dt^2+e^{2Ht}\left(d\rho^2+\rho^2d\Omega^2\right) \ ,
\end{equation}
we make the replacement
\begin{equation}
r \ = \ e^{Ht}\rho
\end{equation}
to yield
\begin{equation}
-dt^2+\left(dr-\frac{r}{L}dt\right)^2+r^2d\Omega^2 \ ,
\end{equation}
where
\begin{equation}
L \ = \ H^{-1} \ .
\end{equation}
Alternatively, one can arrive at this result by starting with the standard form of the de Sitter metric:
\begin{equation}
-\left(1-\frac{r^2}{L^2}\right)dT^2+\frac{dr^2}{1-\frac{r^2}{L^2}}+r^2d\Omega^2
\end{equation}
and performing the analogue of the Gullstrand-Painleve transformation:
\begin{equation}
T \ = \ t \ + \ \int \ dr \ \frac{\frac{r}{L}}{1-\frac{r^2}{L^2}} \ .
\end{equation}
In Cartesian coordinates, this metric has the form
\begin{equation}
-dt^2+\left(dx^i-\frac{r}{L}\frac{x^i}{r}dt\right)^2 \ .
\end{equation}
This is a warp--type metric with
\begin{equation}
N^i \ = \ -\frac{r}{L}\frac{x^i}{r} \ .
\end{equation}
After embedding it becomes
\begin{equation}
N^i \ = \ -(1-f)\frac{x^i}{L}-fv^i \ .
\end{equation}
The energy density is given by
\begin{equation}
\rho \ = \ (1-f)^2\frac{3}{8\pi GL^2}+\frac{f'(1-f)}{4\pi GL}\left(\left(\vec{v}-\frac{\vec{r}}{L}\right)\frac{\vec{r}_s}{r_s}\right)-\frac{f'^2}{32\pi G}\left|\left(\vec{v}-\frac{\vec{r}}{L}\right)\otimes\frac{\vec{r}_s}{r_s}\right|^2 \ .
\end{equation}
Now, if we write
\begin{equation}
\vec{r} \ = \ \vec{r}_0+\vec{r}_s \ ,
\end{equation}
where $\vec{r}_0$ is the position of the center of the bubble relative to the center of the de Sitter universe, and then set
\begin{equation}\label{V2}
\vec{v} \ = \ \frac{\vec{r}_0}{L} \ ,
\end{equation}
the last negative definite term will be \textit{identically} set to zero, and the energy density would be given by
\begin{equation}
\rho \ = \ (1-f)^2\frac{3}{8\pi GL^2}-\frac{r_sf'(1-f)}{4\pi GL^2} \ ,
\end{equation}
which is \textit{positive definite}. Alternatively, this expression may be written in the form:
\begin{equation}
\rho \ = \ \frac{3}{8\pi GL^2}+\frac{1}{8\pi GL^2r^2_s}\left(r_s^3f\left(f-2\right)\right)' \ ,
\end{equation}
where the first term is the background vacuum energy density, and the second is a total divergence that averages to zero upon integration over $d^3r$. This means that the total amount of energy in the universe does not change, and the bubble is effectively just a rearrangement of vacuum energy: we create an underdensity region with $\rho=0$, surrounded by a shell with excess $\rho$.

The mathematical reason for this is that under the condition~\eqref{V2}, $\vec{N}$ would be a gradient:
\begin{equation}
N^i \ = \ \partial_i\phi
\end{equation}
of the function
\begin{equation}
\phi(r) \ = \ -\frac{r^2}{2L} \ + \ \frac{1}{L}\int^{|\vec{r}-\vec{r}_0|}_0 \ dr' \ r'f(r') \ .
\end{equation}
Therefore for a highly specific type of warp drive that moves along the "river flow", it \textit{is} possible to get rid of negative energy. Rather than being "made" of negative energy, this kind of warp bubble may be physically interpreted as a kind of underdensity in vacuum energy. However, if we want the condition~\eqref{V2} to always be valid, we would need the bubble to accelerate. Namely, by integrating~\eqref{V2} from some initial point $r_i$, we obtain
\begin{equation}
r(t) \ = \ r_ie^{Ht} \ , \ v(t) \ = \ \frac{r_i}{L}e^{Ht} \ .
\end{equation}

By inverting the first relation, we find
\begin{equation}\label{Time}
t \ = \ t_H\ln\left(\frac{r}{r_i}\right) \ .
\end{equation}

One can also write $\rho$ as a sum of the "normal" de Sitter energy density and a total 4-divergence that averages to zero:
\begin{equation}
\rho \ = \ \frac{3}{8\pi GL^2} \ + \ \nabla_\mu V^\mu \ ,
\end{equation}
where
\begin{equation}
V^0 \ = \ 0 \ , \ V^i \ = \ \frac{1}{16\pi G}\left(N^i\partial_jN^j-N^j\partial_jN^i\right)-\frac{3}{8\pi GL}\frac{x^i}{L} \ .
\end{equation}
For a metric with only a nontrivial shift vector, the covariant divergence is equal to the "normal" divergence because
\begin{equation}
\nabla_\mu V^\mu \ = \ \frac{1}{\sqrt{-g}}\partial_\mu\left(\sqrt{-g}V^\mu\right) \ ,
\end{equation}
and $g=-1$ for a "shift-only" metric.

Now, let us consider the other components of the SET:
\begin{equation}
T_{ni} \ = \ \frac{1}{16\pi G}\left(\partial_i(\partial_jN^j)-\partial^2N_i\right) \ = \ 0 \ ,
\end{equation}
\begin{eqnarray}
T_{ij} \ = \ \rho\delta_{ij}+\frac{1}{16\pi G}\left(\partial_iN^k\partial_jN^k-\partial_kN_i\partial_kN_j\right)+\\
+\frac{1}{8\pi G}\partial_k\left(N^k\left(K_{ij}-K\delta_{ij}\right)\right)+\frac{1}{8\pi G}\partial_t\left(\delta_{ij}K-K_{ij}\right) \ .
\end{eqnarray}
The second term vanishes identically due to $N^i$ being a gradient, and the third and fourth term can be rewritten as a total divergence. Far away from the bubble, the third term equals
\begin{equation}
-\frac{3}{4\pi GL^2}\delta_{ij} \ ,
\end{equation}
and therefore one can also write $T_{ij}$ as the sum of the "normal" de Sitter value and a 4-divergence that averages to zero:
\begin{equation}
T_{ij} \ = \ -\frac{3}{8\pi GL^2}\delta_{ij}+\nabla_\mu Y^\mu_{ij} \ ,
\end{equation}
with
\begin{eqnarray}
Y^0_{ij} \ = \ \frac{1}{8\pi G}\left(\delta_{ij}\partial_kN_k-\partial_iN_j\right) \ ,\\
Y^k_{ij} \ = \ V^k\delta_{ij}+\frac{N^k}{8\pi G}\left(\partial_iN_j-\partial_kN_k\delta_{ij}\right)+\frac{3}{4\pi GL}\frac{x^k}{L}\delta_{ij}
\end{eqnarray}
This means that the weak energy condition (i. e. energy density measured by a timelike observer moving at velocity $\beta$ in the direction $\vec{n}$) is given by
\begin{equation}
\rho+\beta^2T_{ij}n^in^j \ = \ \left(1-\beta^2\right)\frac{3}{8\pi GL^2}+\nabla_\mu V^\mu+\beta^2n^in^j\nabla_\mu Y^\mu_{ij} \ ,
\end{equation}
i. e. a sum of a positive definite "normal de Sitter" term and the divergence terms that average to zero.

The null energy condition (NEC) is a limiting case of the WEC, obtained in the limit $\beta\rightarrow1$ -- therefore, as the first term vanishes in this limit, the NEC is saturated up to the divergence terms.

\section{Conclusions}\label{sec:conclusions}
In this paper, we studied the model of warp drive embedding into a de Sitter background, and found that it's possible to construct a warp bubble for which:
\begin{itemize}
\item The energy density is strictly non--negative;
\item The weak and null energy conditions are satisfied up to the divergence terms that average to zero.
\end{itemize}
\begin{figure}
    \centering
    \includegraphics[width=0.7\linewidth]{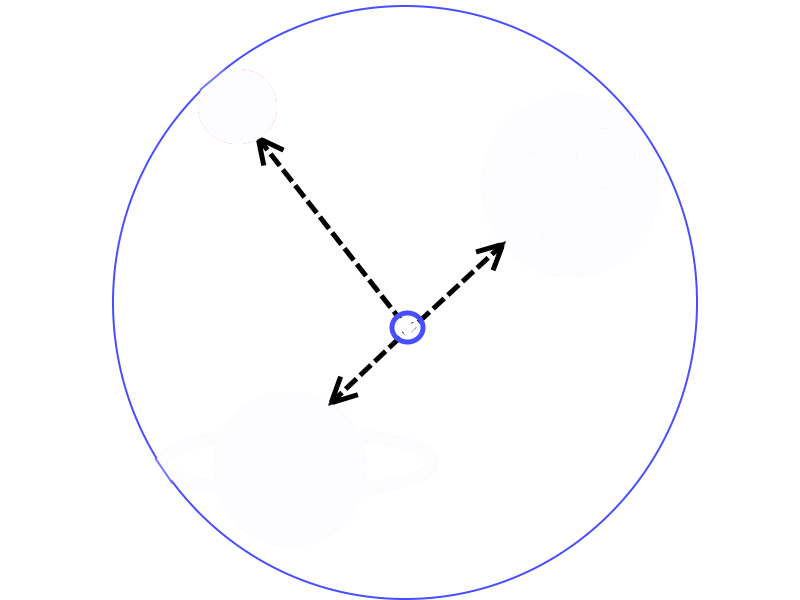}
    \caption{As the warp bubble can only move in the radial direction, one needs to start at the center of the de Sitter universe to be able to choose direction.}
    \label{fig1}
\end{figure}
In order to achieve this, one needs the bubble to move in the radial direction at a velocity exactly equal to the speed of the expansion of the universe. This means that this type of warp drive is not a very practical option for space travel because if you want to choose your direction, you have to start at the center of the universe (otherwise you can only move in one particular direction; see fig.~\ref{fig1}). Besides, one can only realistically traverse distances that are much smaller than the bubble's initial distance from the center of the universe ($r-r_i\ll r_i$); otherwise, according to~\eqref{Time}, the time of travel will be of the order of the Hubble time (which, for our Universe, is about 14.4 billion years).

Nonetheless, this solution is useful as a "proof of concept" because it demonstrates the principal possibility of creating a warp bubble with non--negative energy density through a loophole in the no-go theorem outlined in~\cite{Santiago:2021aup}, namely by considering a spacetime whose shift vector is \textit{both} irrotational (gradient) and asymptotically non-vanishing. Besides, it illustrates the idea of creating a warp drive through a manipulation of dark energy.

It also serves as a mathematical model of vacuum energy underdensities: in principle, it is possible to consider more generic models of these underdensities, such as stationary bubbles, non--spherical bubbles, bubbles in which the energy density is not exactly zero ($f<1$) in the inner region, etc. This research direction may be potentially useful for the general relativistic description of quantum "negative energy" effects (such as Casimir effect, etc.) because those effects represent not actual negative energy but \textit{underdensities} in the background (positive) energy of the quantum vacuum, i. e. suppression of vacuum fluctuations~\cite{Ford:2009vz}. 

This result may also have implications for theories like quintessence~\cite{Ratra:1987rm,Caldwell:1997ii,Tsujikawa:2013fta}, "dark fluid"~\cite{Arbey:2005fn,Arbey:2006it}, and superfluid vacuum~\cite{Dirac:1951,Dirac:1952,Sinha:1976dw,Sudarshan:1976cb,Sinha:1978gu} that model dark energy as a kind of fluid. Analogue gravity setups mimicking a de Sitter universe are also known to exist, so it may be possible to create this kind of bubbles and study their effects within these analogue systems~\cite{Barcelo:2005fc,Fischer:2002jn,Fedichev:2003id,Fedichev:2003bv,Fischer:2004bf,Weinfurtner:2004mu,Nambu:2023tpg}.

\section{Declaration of Competing Interest}
The authors declare the following financial interests/personal relationships which may be considered as potential competing interests: Remo Garattini reports financial support was provided by Limitless Space Institute. Kirill Zatrimaylov reports financial support was provided by Limitless Space Institute. Remo Garattini reports a relationship with Limitless Space Institute that includes: funding grants. Kirill Zatrimaylov reports a relationship with Limitless Space Institute that includes: funding grants. If there are other authors, they declare that they have no known competing financial interests or personal relationships that could have appeared to influence the work reported in this paper.

\section{Acknowledgements}
We are grateful to Harold "Sonny" White for his support, and to Rene Meyer for the discussions and for the suggestion to look into warp bubbles in the context of the de Sitter and anti-de Sitter spacetimes. The work is supported by the 2023 LSI grant "Traversable Wormholes: A Road to Interstellar Exploration".

\end{document}